# Mexican hat-like valence band dispersion and quantum confinement in rhombohedral ferroelectric α-In$_2$Se$_3$

Geoffroy Kremer[1], Aymen Mahmoudi[2], Meryem Bouaziz[2], Mehrdad Rahimi[3], François Bertran[4], Jean-Francois Dayen[5,6] Maria Luisa Della Rocca[3], Marco Pala[7], Ahmed Naitabdi[8], Julien Chaste[2], Fabrice Oehler[2], and Abdelkarim Ouerghi[2]

[1] Institut Jean Lamour, UMR 7198, CNRS-Université de Lorraine, Campus ARTEM, 2 allée André Guinier, BP 50840, 54011 Nancy, France
[2] Université Paris-Saclay, CNRS, Centre de Nanosciences et de Nanotechnologies, 91120, Palaiseau, France
[3] Laboratoire Matériaux et Phénomènes Quantiques, Université Paris Cité, CNRS UMR 7162, F-75013 Paris, France
[4] Synchrotron SOLEIL, L'Orme des Merisiers, Départementale 128, 91190 Saint-Aubin, France
[5] Université de Strasbourg, IPCMS-CNRS UMR 7504, 23 Rue du Loess, 67034 Strasbourg, France
[6] Institut Universitaire de France, 1 rue Descartes, 75231 Paris cedex 05, France
[7] DPIA, University of Udine, 33100 Udine, Italy
[8] Laboratoire de Chimie Physique – Matière et Rayonnement, UMR 7614, Sorbonne Université, 4 Place Jussieu, 75005 Paris, France

Two-dimensional (2D) ferroelectric (FE) materials offer a large variety of electronic properties depending on chemical composition, number of layers and stacking-order. Among them, α-In$_2$Se$_3$ has attracted much attention due to the promise of outstanding electronic properties, attractive quantum physics, in- and out-of-plane ferroelectricity and high photo-response. Precise experimental determination of the electronic structure of rhombohedral (3R) α-In$_2$Se$_3$ is needed for a better understanding of potential properties and device applications. Here, combining angle resolved photoemission spectroscopy (ARPES) and density functional theory (DFT) calculations, we demonstrate that 3R α-In$_2$Se$_3$ phase exhibits a robust inversion of the valence band parabolicity at the Γ point forming a bow-shaped dispersion with a depth of $140 \pm 10$ meV between the valence band maximum (VBM) along the ΓK direction of the Brillouin zone (BZ). Moreover, we unveil an indirect band gap of about 1.25 eV, as well as a highly electron doping of approximatively $5 \times 10^{12}$ electrons/cm² at the surface. This leads to surface band bending and the formation of a prominent electron accumulation layer. These findings allow a deeper understanding of the rhombohedral α-In$_2$Se$_3$ electronic properties underlying the potential of III−VI semiconductors for electronic and photonic technologies.

Two dimensional (2D) van der Waals (vdW) III−VI semiconductors have drawn intense attention due to their unique electronic properties [1], [2], [3], [4], [5]. Among these materials, $In_2Se_3$ holds different phases and several polytypes, with potential applications in various domains, going from electronics, photonics to thermoelectricity, due to its good electron mobility, excellent photoresponse, exotic ferroelectricity [6], and unique electronic band structure [7]. $In_2Se_3$ has been extensively investigated in different forms, such as thin films, nanocrystals, and nanostructures [6]. In particular, the investigations of III−VI semiconductors has propelled the study of post-transition metal two-dimensional (2D) semiconductors [8], [9], [10]. When engineering into real devices, $In_2Se_3$ electronic properties may be further modified via chemical doping [11]. Moreover, it has been also integrated in van der Waals heterostructures with other 2D compounds, allowing to further tune electronic, magnetic and optical properties of the 2D assembly or to perform band structure engineering [6]. Therefore, some key aspects of 2D material technological integration directly relate to fundamental surface and interface physics, contacts, and energy band profile control. As a consequence, before transferring from science to technology [12], it is crucial to understand the fundamental properties of III-VI semiconductors, and to investigate the change of their electronic properties, particularly of the band structure (up to room temperature) [13], with the crystalline phase [3].

2H stacking is generally more stable than 3R stacking in few layers graphene [14], [15] and in many van der Waals (vdW) layered transition metal dichalcogenide (TMD) materials [16–18]. The ferroelectric behavior of 3R-stacked $MX_2$ results from interlayer charge transfer between layer structures lacking inverse symmetry. However, the impact of intralayer (ionic displacement) and interlayer (charge transfer) effects on ferroelectricity remains largely unexplored in spontaneous ferroelectric van der Waals materials [19], [20]. Remarkably, α-$In_2Se_3$ in its hexagonal structure (2H) hosts a fascinating combination of a robust *n*-type doping and a peculiar band structure, leading to the formation of a quantum confined electron gas (2DEG) directly at its surface [7], [21]. The 2H α-$In_2Se_3$ 2DEG maintains considerably high electron density even at room temperature, comparable to conventional AlGaN/GaN systems, enabling the observations of anomalous optical response and quantum Hall effect [22]. The 2DEG in 2H α-$In_2Se_3$ represents an alternative platform for achieving high performance devices not subjected to the constraints of stable stoichiometry and alloy composition [19] required for AlGaN/GaN.

Besides the 2H-hexagonal stacking, the α phase of $In_2Se_3$ possess a 3R stacking form, where the basic block of Se-In-Se-In-Se atoms sequence stacks vertically in a rhombohedral structure. It has been recently demonstrated that the ferroelectric properties of α-$In_2Se_3$ are highly dependent on its stacking form. A distinct ferroelectric polarization switching mechanisms in metal/α-$In_2Se_3$/metal heterostructures has been measured depending on the nature of the embedded polytype [19]. The 2H and 3R structures show a stacking-dependent ferroelectric domain walls dynamic, implying larger hysteresis windows for the 3R case, and different intralayer dynamics related to the ferro-paraelectric phase transition induced by high electric fields[25]. Thus, exploring and controlling the stacking dependent physical properties in two-dimensional materials is of fundamental interest, opening new routes for interlayer interactions engineering. While fundamental understanding of 2H α-$In_2Se_3$ electronic structure has recently advanced, the 3R $In_2Se_3$ polytypes lacks of such an equivalent knowledge and still requires detailed investigations. Experimental band structure mapping using angle resolved photo-electron spectroscopy (ARPES), associated to density functional theory (DFT) simulations have the potential to answer this question.

In this work, we report the electronic properties of 3R-α-In$_2$Se$_3$. Complementary micro- Raman spectroscopy analysis and scanning transmission electron microscopy (STEM) was conducted to study the structural properties of the crystal exploring the vibration frequencies of phonons corresponding to the characteristic vibrational modes of the sample. ARPES was used to investigate the 3R α In$_2$Se$_3$ band structure, electronic band gap and the position of the valence band maximum relative to the Fermi level. Such result is further supported by the excellent agreement with the density functional theory (DFT) calculated band structure with the experimental data. Our results show an indirect band gap of ~1.25 eV, with the bottom of the conduction band localized at the center of the Brillouin zone, just below the Fermi level. Furthermore, we demonstrate the presence of mexican hat-like valence band dispersion in 3R α In$_2$Se$_3$ at the Γ point using ARPES measurements and DFT calculations. Such a peculiar electronic structure results in density of state (DOS) singularities opening exciting opportunities for engineering novel electronic devices and for exploring new material solutions with enhanced thermoelectric properties. We show that thin flakes exfoliated from the same 3R α In$_2$Se$_3$ crystals display ferroelectric (FE) properties when integrated into FET-like devices. By performing thermoelectric experiments, we confirm a *n*-type doping of the flakes with a large Seebeck coefficient of the order of 200μV/K in absolute value. The extracted electron density of approximatively $5 \times 10^{12}$ electrons/cm² matches correctly the value determined from ARPES and DFT calculation. The ensemble of these observations confirms the great potential of α In$_2$Se$_3$ for future applications where stacking engineering can play a major role.

**Structural and chemical properties**

III–VI materials crystallize in a common layered van der Waals configuration, with well-defined covalently-bond monolayers in the plane, weakly coupled to each other along the out-of-plane axis by van der Waals forces (Figure 1(a)). For α-In$_2$Se$_3$ each monolayer consists of two different indium atoms with different (covalent) coordination polyhedra: In[1] is coordinated tetrahedrally by four Se atoms while In[2] is octahedrally coordinated by six Se [1]. The inner structure of an individual monolayer thus decomposes into five sub-layers, in a Se-In[1]-Se-In[2]-Se repetition along the vertical direction. Figure 1(a) shows a schematic of 3 vertical stackings of such a quintuple sub-layer (QL) unit. In addition, α-In$_2$Se$_3$ possess several polytypes depending on the relative position and in-plane orientation of each QL. In Figure 1(a), the 3 QLs show a small in plane translation, implying a ABC stacking order characteristic of the rhombohedral configuration of the 3R α-In$_2$Se$_3$ polytype. Figure 1(b) shows a schematic of the bulk and surface-projected hexagonal Brillouin zone of α -phase In$_2$Se$_3$. The crystal unit cell being hexagonal, the important high symmetry points are Γ, M and K in the three-dimensional bulk Brillouin zone (3D BZ), and their respective planar projection (2D BZ). In Figure 1(c), we show the electronic structure of bulk 3R α-In$_2$Se$_3$, as calculated from the Heyd-Scuseria-Ernzerhof (HSE) hybrid functionals [23]. The minimum of the conduction band appears in the Γ point, while the valence-band maximum (VBM) is located near the Γ point, thus resulting in a negative effective mass for the highest valence band at the Γ point. Most interestingly, the energy difference between the VBM and the highest valence band state near the Γ is higher, with the Mexican-hat-like landscape extending over a significant fraction of the Brillouin zone. This unusual character of the band structure gives rise to a high DOS and an almost one-dimensional (1D) like van Hove singularity near the VBM [24]. In general, a large near the Fermi level ($E_F$), in general, would lead to instabilities and transitions to different phases such as magnetism, superconductivity, and other phenomena [25]. Also, sharp features in the DOS near $E_F$ are often indicative of a material with a large thermoelectric Seebeck coefficient [26].

The specific crystalline phase of In₂Se₃ is generally difficult to identify. There are several different phases (α, β, γ, δ and κ), some of which exhibit similar properties. For example, the γ- and δ- phases have hexagonal and trigonal crystal structures respectively, whereas the α- and β- phases have highly similar rhombohedral structures. In order to probe the chemical and structural properties of 3R α-In$_2$Se$_3$ single crystal (HQ Graphene), micro-Raman and high-resolution x-ray photoemission spectroscopy investigations were performed on the same sample. The 3R α-In$_2$Se$_3$ micro-Raman spectrum measured at room temperature is reported in Figure 1(d). We can distinguish from this spectrum the typical vibrational modes reported previously for bulk α-In$_2$Se$_3$ [6], [13]. As shown in Figure 1(d), multiple micro-Raman peaks are observed near 88, 104, 159, 179, 193 and 250 cm$^{-1}$ in 3R α-In$_2$Se$_3$ crystals, which can be assigned to the E and A modes of α-In$_2$Se$_3$, respectively [2], [27], [1]. We can see that the E mode of 3R α-In2Se3 is stronger than that of E mode. The intensity ratio between E and A modes can therefore be used to identify the 3R α-In$_2$Se$_3$ polytypes. high-resolution x-ray photoemission spectroscopy (XPS) measurements were carried out on the 3R α-In$_2$Se$_3$ crystals at the Cassiopée beamline of the synchrotron Soleil (France) [28], [29], [30]. High resolution XPS spectra for In (In 4d), and Se (Se 3d) are recorded at 110 eV (Figure 1 (f), and (e), respectively). The In spectrum presents two doublets, which can be attributed to the In 4d$_{5/2}$ and In 4d$_{3/2}$ with a spin-orbit (SO) splitting of 0.8 eV. The high intensity doublet corresponds to the In atoms embedded in the 3R α-In$_2$Se$_3$, while the second In 4d at higher binding energy is the signature of a defective/sub-stoichiometric 3R α-In$_2$Se$_3$ with Se vacancy or In$_2$O$_3$ [31]. Similarly, two components are also present for the Se 3d peak, corresponding to the Se 3d$_{3/2}$ and 3d$_{1/2}$ at 54.4 eV and 55.2 eV, respectively. The obtained binding energy values agree with previously reported values [6], [32]. The crystalline phase of the sample was demonstrated by energy-dispersive X-ray spectroscopy (EDX) and X-ray photoelectron spectroscopy (XPS), which mapped the elemental distribution of Se and In in a selected area of the In$_2$Se$_3$. The EDX images (Figure S2) and XPS spectra demonstrate a continuous film in which the atomic range of Se and In could be identified. The intensity of the In and Se maps across the sample also indicates that the Se defect stoichiometry is lower than 1%.

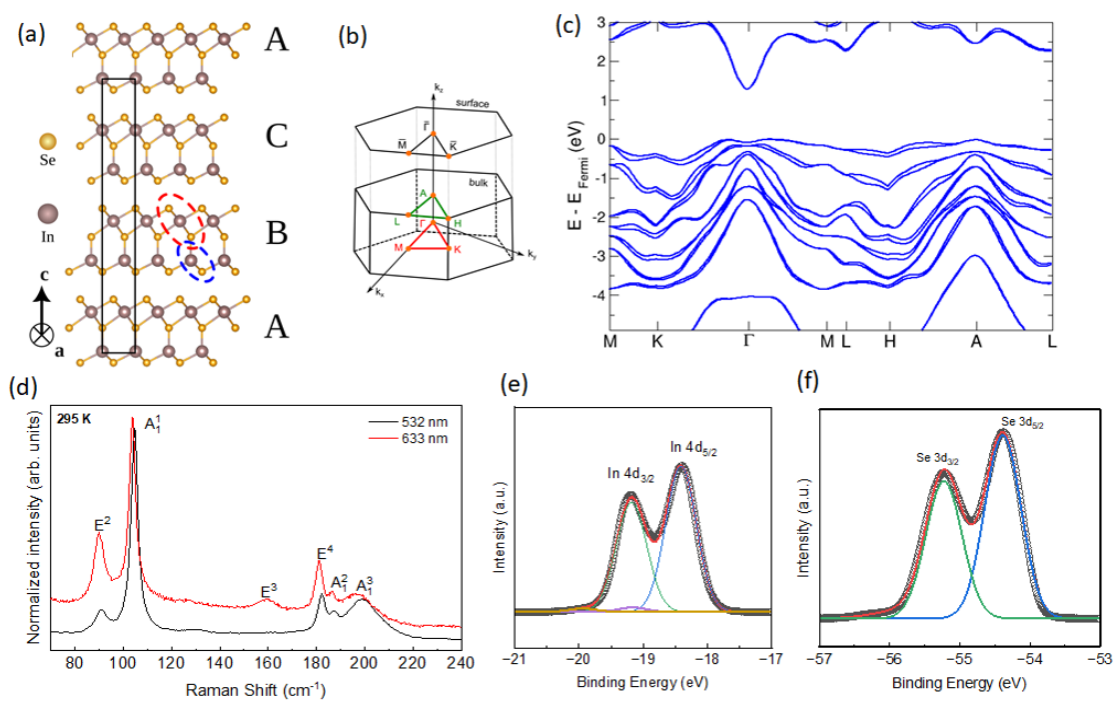

**Figure 1:** (a) Side view of the surface crystal structure of 3R α-In$_2$Se$_3$. The black line shows the conventional unit cell. (b) Corresponding three-dimensional (3D) Brillouin zone (BZ) and its 2D projection in the (001) plane. (c) Calculated band structure. (d) Micro-Raman spectroscopy spectra for 3R α-In$_2$Se$_3$ crystals using 532 and 633 nm laser. (e, f) XPS of 3R α-In$_2$Se$_3$ at hν = 110 eV: In 4d and Se 3d.

In order to confirm the nature of our polytype, a part of our commercial 3R α-In$_2$Se$_3$ bulk crystal was analyzed by Transmission Electron Microcopy (Titan THEMIS, Thermofisher). Figure 2(a) shows a corresponding low-resolution scanning-TEM (STEM) image. The stacking structure along the vertical axis is already apparent. For 3R α-In$_2$Se$_3$ (symmetry R3m) [1], all atoms in the unit cell can be grouped in triplets (In$_1$, In$_2$, Se$_1$, Se$_2$, Se$_3$) which are all located in Wickhoff position *3a*. Each triplet is comprised of equivalent positions (0, 0, z) (2/3, 1/3, 1/3+z) (1/3, 2/3, 2/3+z) with arbitrary z. Consequently, there are many systematic extinctions in the structure factor and HKL reflections are only allowed when H-K-L=3n, with n an arbitrary integer. We retrieve this particular rule in the experimental diffraction pattern presented in Figure 2(b). We also determine an experimental ratio c/a=0.1408 for the crystal. The matching c/a ratio and the observed systematic extinction pattern are a strong indication for the 3R α-In$_2$Se$_3$ nature of our sample [1]. To confirm the stacking order of the sample, we show in Figure 2(c) an atomically-resolved STEM image of the In$_2$Se$_3$ crystal. We distinguish easily the van der Waals gap between each consecutive lamella along the vertical direction. Inside each lamella, we observe a clear asymmetry between the bottom In1 atom coordinated tetrahedrally by four Se atoms (blue ellipse) and the top In2 atom coordinated octahedrally by six Se atoms (red ellipse). We clearly recognize the α-In$_2$Se$_3$ monolayer structure (see also Figure 1(a)). As there is no rotation between consecutive layers, but a simple in-plane translation with a period of three layers along the vertical direction [1], we confirm the polytype agrees with the 3R α-In$_2$Se$_3$ configuration.

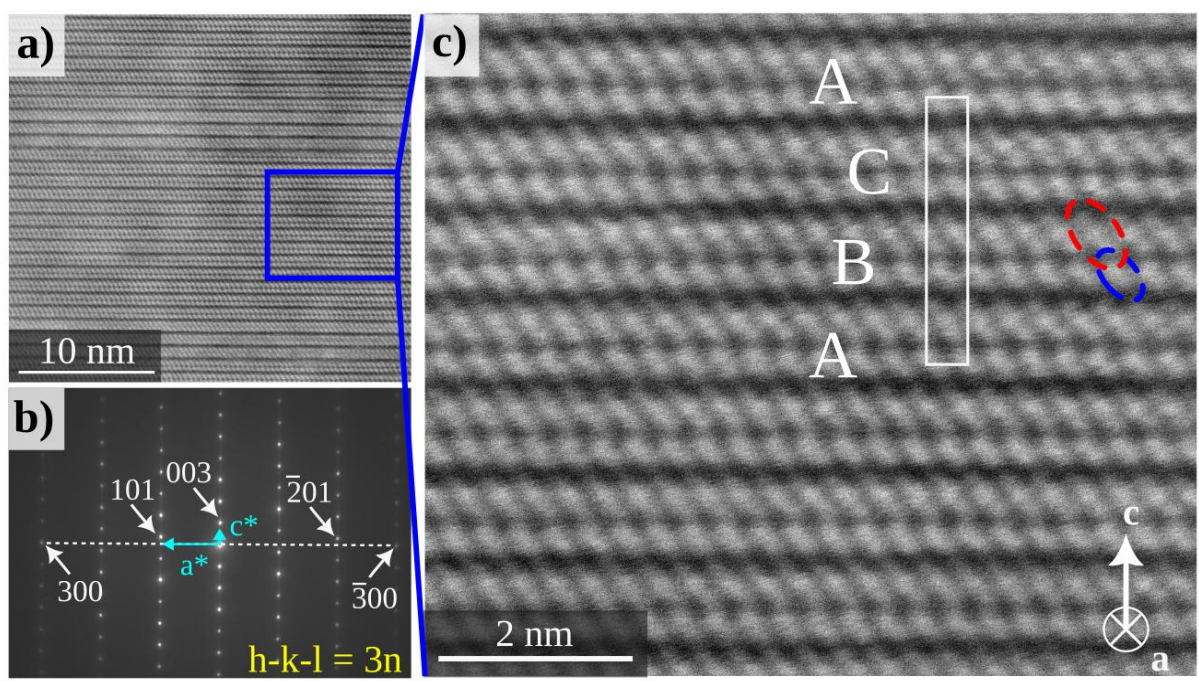

**Figure 2:** Transmission Electron Microscopy (TEM) characterization of the In$_2$Se$_3$: (a) Large Bright-field scanning transmission electron microscopy (STEM) image of 3R α-In$_2$Se$_3$, (b) Selective area diffraction patterns of In$_2$Se$_3$ (c) high resolution STEM image of 3R α-In$_2$Se$_3$.

**Electronic band structure: ARPES vs DFT**

We now focus on the electronic properties of 3R α-In$_2$Se$_3$ as measured by ARPES. First, we show in Figure 3(a-c) a series of isoenergetic contours in the ($k_x$, $k_y$) plane obtained between BE=0 eV (E$_F$) and BE=6 eV at a photon energy of hv = 90 eV. The use of this photon energy is sufficient to reach the Brillouin zone (BZ) edges. In particular, we observe in Figure 3(a) the evidence of the periodicity of the electronic band structure at this energy between the $\bar{\Gamma}$ point of the first Brillouin zone (BZ), and the $\bar{\Gamma}$ points of the second BZ. At this energy, a single intensity point is observed at the $\bar{\Gamma}$ points and no photoemission intensity at other places in the BZ. By superimposing the projected 2D hexagonal BZ (red line) on the isoenergetic contour, we show an excellent agreement using the in-plane lattice parameter of the crystal (a= 4.026 Å) [1]. We have consequently located the other high-symmetry points of the BZ, namely $\bar{\Gamma}$, $\bar{K}$ and $\bar{M}$. Other isoenergetics contours are plotted in Figure 3(b, c) corresponding to BE equals to 1.7 and 6 eV respectively. We experimentally resolve a lack of photoemission intensity between approximatively 1.7 eV BE (Figure 3(b)) and E$_F$ (Figure 3(a)) which we ascribe to the electronic band gap of the material, reflecting its semiconducting character. At -6 eV BE (Figure 3(c)), we can resolve some spectral weight at the $\bar{K}$ point, no photoemission intensity at the $\bar{M}$ point and a circular structure around the $\bar{\Gamma}$ point. These behaviors are very similar to what we observed for the 2H stacking[1], and we did not resolve at this point different spectroscopic signatures between the 3R and 2H α-In$_2$Se$_3$.

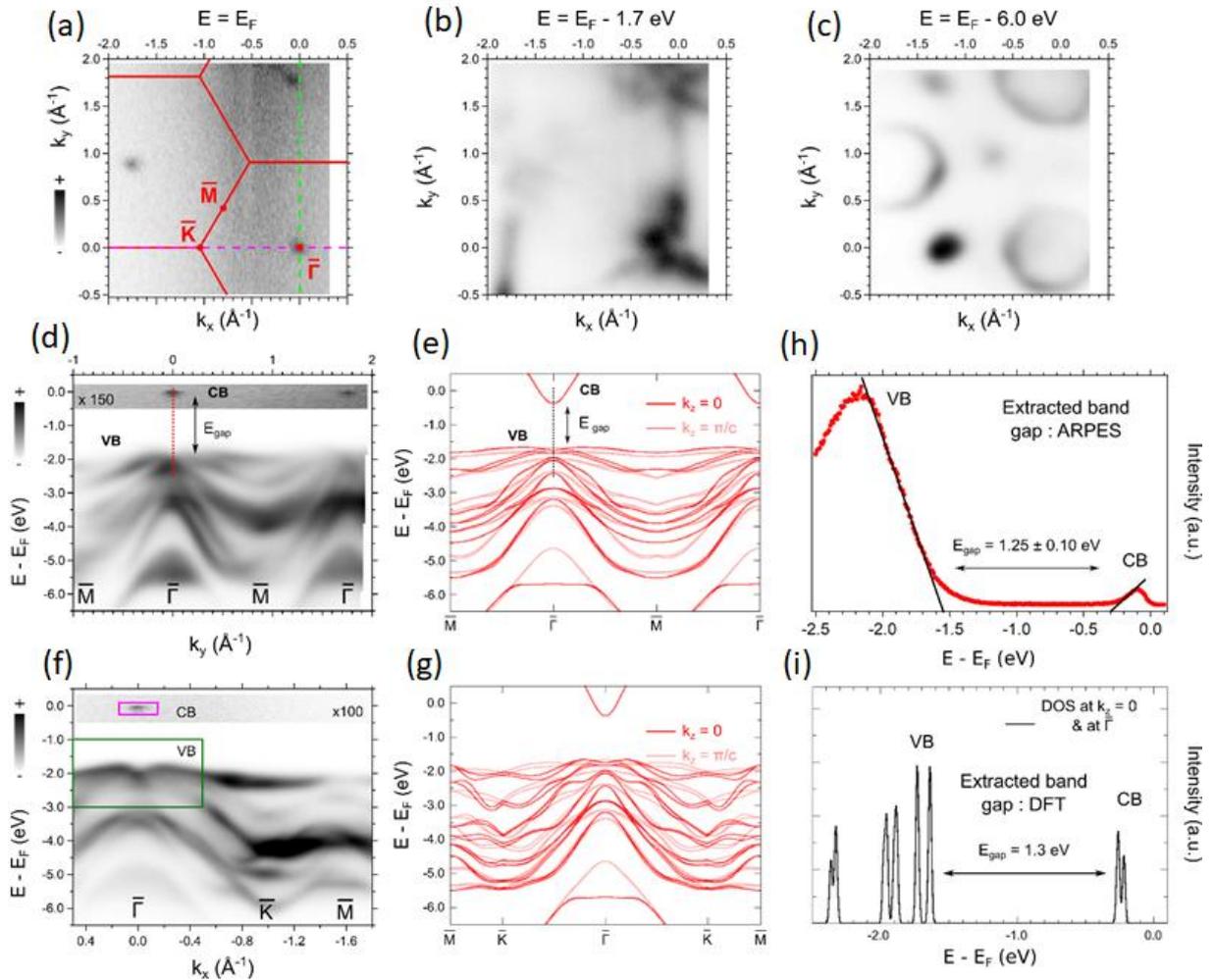

**Figure 3:** Constant ARPES energy surfaces for 3R α-In$_2$Se$_3$ (hν = 90 eV) with 2D projections taken at E - E$_F$ equals (a) 0 eV (b) - 1.7 eV and (c) - 6 eV. In panel (a), red line corresponds to the 2D hexagonal Brillouin zone (BZ) obtained using the in-plane cell parameter of α-In$_2$Se$_3$ (a = 4.0 Å). Corresponding ARPES spectra along (d) $\bar{M} - \bar{\Gamma} - \bar{M} - \bar{\Gamma}$ and (f) $\bar{\Gamma} - \bar{K} - \bar{M}$ high-symmetry directions corresponding to green and pink dashed lines in panel (a), respectively. (e, g) Corresponding DFT calculations with HSE functional for two distinct values of k$_z$ in the 3D BZ, namely k$_z$ = 0 (dark red) and k$_z$ = π/c (light red). A rigid energy shift of +1 eV has been applied to fit the ARPES dispersion. (h) Normal emission energy distribution curve extracted from ARPES (red dashed line in panel (d)) showing the amplitude of the band gap. (i) Corresponding value extracted from density of state (DOS, as extracted from DFT calculations at k$_z$ = 0 and at the center of the 2D BZ (black dashed lines in panel (e)).

To go in deeper details, we plot in Figure 3 (d, f) the ARPES dispersion along the $\bar{M} - \bar{\Gamma} - \bar{M} - \bar{\Gamma}$ and $\bar{\Gamma} - \bar{K} - \bar{M}$ high-symmetry directions of the BZ, respectively. These directions correspond to the green and pink dashed lines in Figure 3(a). We directly visualize the semiconducting character of 3R α-In$_2$Se$_3$ with the bottom of the conduction band and the top of the valence band localized close to the center of the BZ. The Fermi level in 3R α-In$_2$Se$_3$, is ~150 ± 15 meV above the conduction band minimum, resulting in intrinsic *n*-type charge carrier doping of the "as grown" material. By comparing our ARPES results to DFT calculations, we find an excellent agreement of the electronic dispersion along both high symmetry directions, except for the complex variation of photoemission spectral weight which cannot be captured by this type of calculation. The DFT-calculated dispersions are shown in Figure 3(e, g) for two values of k$_z$ in the 3D BZ, namely k$_z$ = 0 (dark red) and k$_z$ = π/c (light red), along the $\bar{M} - \bar{\Gamma} - \bar{M} - \bar{\Gamma}$ (Figure 3(e)) and $\bar{\Gamma} - \bar{K} - \bar{M}$ directions (Figure 3(g)). From this comparison between ARPES and DFT, it is difficult to disambiguate between these two k$_z$ values in the 3D BZ, i.e. what is the vertical momentum probed using this particular photon energy (hν = 90 eV). Nevertheless, by looking at the dispersion of the CB, we are likely probing the band structure at the $\bar{\Gamma}$ point (see Figure S1 in the appendix for larger energy scale DFT calculations). A striking point of this comparison is the dispersion of the band at -5.5 eV at $\bar{\Gamma}$ as measured by ARPES. Experimentally, we observe a broadening that we can ascribe to the continuous superposition of the bands (with different k$_z$) between the two high symmetry points in the 3D BZ, i.e., at k$_z$ = 0 and k$_z$ = π/c, as illustrated by the superposition of the DFT electronic dispersions at these two points. This effect is known as k$_z$ broadening and is typical of quasi 2D systems [32]. It is particularly pronounced for electronic states with out-of-plane character such as bands with p$_z$ derived state, as we have discussed in our recent publication on 2H α-In$_2$Se$_3$[4]. Another noticeable agreement between ARPES and DFT is the amplitude of the band gap, that we determine experimentally equal to 1.25 ± 0.1 eV with ARPES in Figure 3(h) by plotting a normal emission energy distribution curve, and theoretically equal to 1.3 eV using HSE hybrid functional as extracted from the DOS in Figure 3 (i). Typically, HSE-type functionals are known to give quantitative estimates of the band gap of semiconducting materials, which is also in good agreement with the GW calculations [33] or photoluminescence measurements [2].

**Inverted valence band parabolicity and 2D electron gas formation**

Figure 3 (f) shows a large scale ARPES scan along the $\bar{\Gamma} - \bar{K} - \bar{M}$ direction for which we focus on two pertinent regions in energy space, namely green rectangle in the vicinity of the VB maximum and pink rectangle close to the Fermi energy, both close to the center of the BZ. In Figure 4 we present further ARPES data at hν = 108 eV for which the out-of-plane momentum is again close to 0 i.e. to the Γ point in the 3D BZ. The ARPES spectra close to the VB maximum reveal a parabolicity inversion around Γ, resulting in a Mexican-hat-like energy surface around the Γ point. The change of the sign of the HVB parabolicity can be explained by two simultaneous effects: the strong interactions with the many underlying valence bands, which tend to decrease the HVB effective mass, and the negligible interaction with the upper conduction band, which would otherwise increase it [34]. Figure 4(b) shows the corresponding energy distribution curves (EDCs) taken at different values of $k_x$. The experimental data, allowing to determine the position of the VBM and the depth of the local minimum arising from the band parabolicity inversion are Δk=0.22 ± 0.1 Å and ΔE=140 ± 10 meV, respectively. These values agree within the uncertainty of our DFT predictions.

A more detailed inspection of the CB dispersion we measured in panel (a), and previously observed at hν = 90 eV in Figure 3, is presented in Figure 4(c). It shows that the CB consists of a single dispersive parabolic band. We noticed that the sample revealed lower doping inhomogeneity on the surface. DFT calculations show that the conduction band structure differs in the KGK and HAH directions. However, ARPES measurements reveal the same structure in both directions. Based on these observations, we can attribute this band to the surface states [7]. This feature is typical of surface charge accumulation layers, which is a key-signature of quantum confinement and it is interpreted as the fingerprint of the occurrence of a 2D electron gas (2DEG). The formation of the 2DEG is due to a surface band bending and the formation of a prominent electron accumulation layer. The origin of the band bending is the presence of donor surface states that can donate their electrons to the conduction band of the $In_2Se_3$, therefore inducing a significant electron accumulation at the surface and a consequent downward band bending. Based the XPS measurement the doping can be attributed to the intrinsic Se vacancies. As in the 2H stacking case, the experimentally observed position of $E_F$ inside the CB can be explained by a strong electron doping of the crystal[4]. We can quantify the dispersive nature of these specific band using a parabolic dispersion. We obtain an effective mass (m*) of 0.1 $m_0$ with energy minimum located at 110 ± 15 meV BE. The discrepancy in energy position observed in the parabolic dispersion using ARPES at 90 and 108 eV is attributed to inhomogeneous doping on the surface. This 2DEG exhibits one confined state which corresponds to an electron density of $5 \times 10^{12}$ electrons/cm² using the 2D Luttinger volume $n_{2D} = k_F^2/2\pi$ formula, also confirmed by thermoelectric measurements (see next discussion). As shown in Figure 4(d), we also performed a 1D self-consistent Poisson-Schrodinger simulation of the potential well and of the accumulation charge at the surface of 3R α-$In_2Se_3$. To fit the experiments, we assumed an effective mass of 0.1 $m_0$, a relative dielectric permittivity of 17 $\varepsilon_0$, a temperature of 300 K and a donor sheet density of $4 \times 10^{13}$ cm$^{-2}$ distributed along the first 3 nm. We found only one confined state below the Fermi level at the energy of $E_1$ = -110 meV and thus contributing to an electron density of about $5 \times 10^{12}$ cm$^{-2}$, which remains compatible with the ARPES data. For clarity, the shape of the absolute value of this confined wavefunction is also plotted.

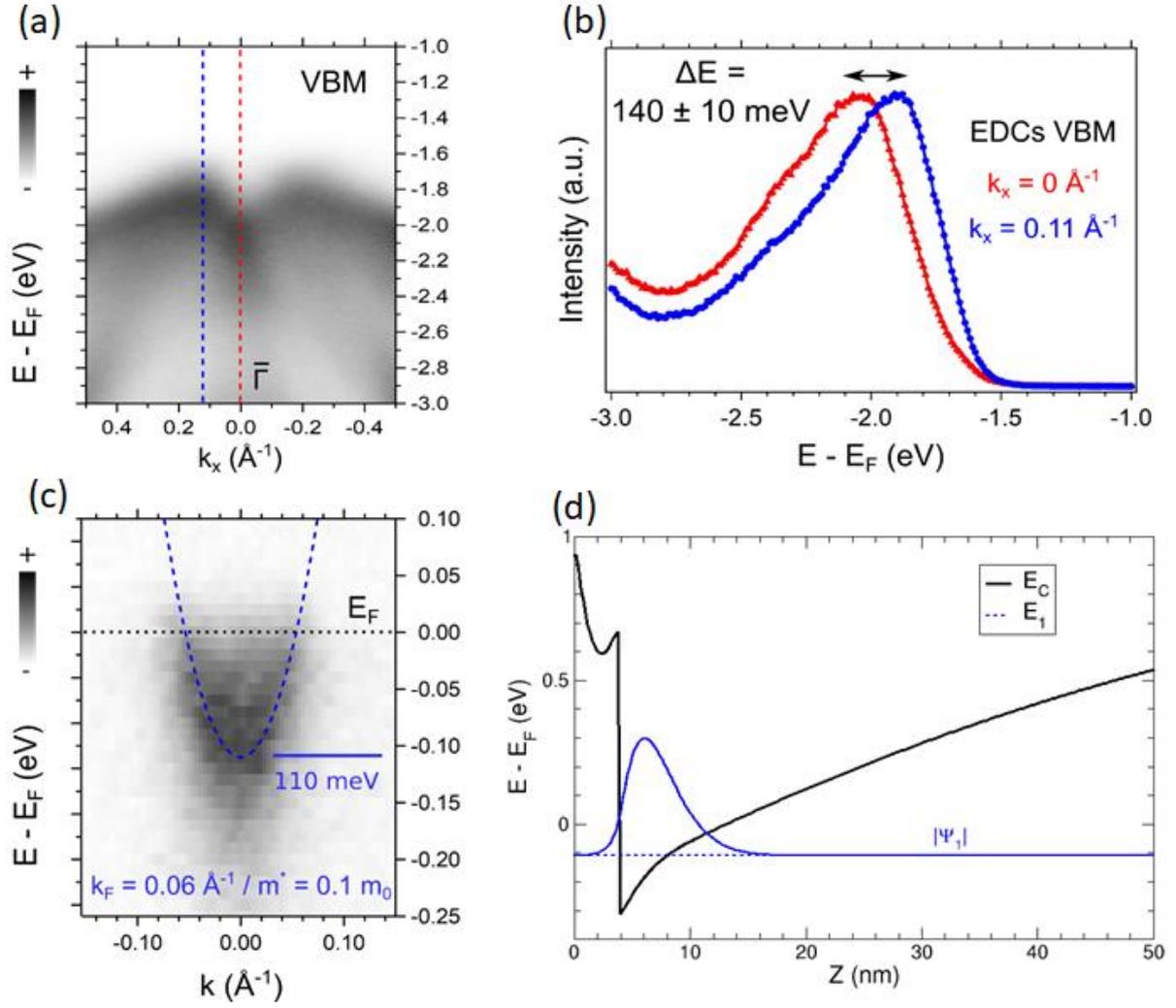

**Figure 4:** ARPES spectra for 3R α-In$_2$Se$_3$ (hv = 108 eV) along (a) the $\bar{\Gamma} - \bar{K} - \bar{M}$ high-symmetry direction. It corresponds to the zoom-in close to the VB maximum as represented by the green rectangle in Figure 3(f). It shows an inversion of the band parabolicity. (b) Corresponding energy distribution curves (EDCs) taken at different values of k$_x$ and showing a 140 meV energy difference between the VB maximum and its position at the center of the BZ. (c) ARPES spectrum obtained in the vicinity of the Fermi energy as represented by the pink rectangle in Figure 3(f) (slightly different spatial region on the sample), showing one parabolic band associated to a two-dimensional electron gas (2DEG). Blue dashed line corresponds to a fit with a nearly free electron dispersion. (d) Conduction band profile along the confinement direction (black line) and first wave function (blue line) extracted using the QW model and experimental parameters obtained from (c) (E$_c$: Conduction band, E$_1$: quantum well state, Z(nm): the thickness of the sample).

**Transport properties**

We have fabricated FET-like devices based on multilayer 3R α-In$_2$Se$_3$ flakes exfoliated from the same single crystals used for the ARPES investigation. The 3R α-In$_2$Se$_3$ flake (~100 nm thick) is coupled to an Au local gate through an hBN flake (~30-40 nm thick), used as a dielectric spacer. Source and drain electrodes are fabricated on the hBN flake by standard microfabrication and the 3R α-In$_2$Se$_3$ flake is dry transferred over it just after mechanical exfoliation. As detailed in ref [35], this fabrication approach allows probing the 3R α-In$_2$Se$_3$ flake surface in contact with the electrodes that remains fresh and protected from oxidization under air exposure. Figure 5(a) shows an atomic force microscopy (AFM) image of a representative 3R α-In$_2$Se$_3$-based FET, an optical image of the same device is also shown in Figure 5(d) allowing to better visualize its detailed structure. The thicknesses of the 3R α-In$_2$Se$_3$ and hBN flakes are typically evaluated by averaging 50 lines profiles taken on zoomed parts of the AFM images (Figure 5(b) and (c)).

The electric and thermoelectric characterizations of the 3R α-In$_2$Se$_3$-based FET are performed under high vacuum (~10$^{-7}$ mbar) at 35°C. Figure 5(f) shows the 2-point I-V measurements obtained by voltage biasing the device in Figure 5(a, d) at different temperatures, from 35°C to 200°C. The source-drain voltage is swept first from negative to positive values (solid dots) and then from positive to negative values (open dots). A non-linear and non-symmetric I-V trend is visible, which can be attributed to the different injection barriers at the source and drain contacts. Moreover, the I-V measurement shows an evident hysteretic behavior, with a current jump around ±1.5 V and a small switching ON/OFF ratio. The I-V hysteresis in the 2-point measurement is characteristic of memristive behavior and can be explained by the asymmetric modulation of the Schottky barriers by residual in plane ferroelectric (FE) polarization. Following the work of Xue et al., residual polarization at the contact interface causes either negative or positive charge accumulation, thereby inducing a change in the effective Schottky barrier. The difference in the measured current at the same VDS can be attributed to different residual ferroelectric polarization depending on the poling history [36]. Figure 5(f) show that the observed hysteresis is robust with the temperature, at least up to 200 °C (curves are vertically shifted for the sake of clarity). Note that at temperature higher than 200 °C, a phase transition from the α to the β phase could occur, so we generally avoid temperature higher than that during the electrical characterization[21].

The thermoelectric analysis is performed soon after the electrical characterization of the devices by a DC approach. A DC current is applied to a micro-heater to create a temperature gradient in the longitudinal direction of the 3R α-In$_2$Se$_3$ flakes. Such a gradient is measured by monitoring the 4-point resistance of the source and drain electrodes that, shaped in a nanowire geometry, act simultaneously also as local thermometers. The thermoelectric voltage between these two electrodes is then measured using a Keithley 2182A nanovoltmeter and the resulting Seebeck coefficient extracted. An example of Seebeck coefficient measured as a function of the gate voltage of another device is shown in Figure 5(f), analogous results have been obtained on a total of 6 devices. The 3R α-In$_2$Se$_3$ flake shows systematically a noisy Seebeck signal (red data) with an average absolute value in the 100-200 µV/K range (black data), depending on the measured device [35]. These values are in good agreement with theoretical calculations for the case on monolayer α-In$_2$Se$_3$ [37]. This noisy character is possibly related to a robust in-plane residual polarization in the flake induced by the applied source–drain voltage[26]. The Seebeck coefficient is negative and shows no dependence on the gate voltage in the whole explored range. Its negative sign indicates that electrons are the majority charge carriers with a charge density extracted by the Mott formula equal to $10^{12}$-$10^{13}$ elec/cm$^2$. All these conclusions agree well with the ARPES measurements, revealing the bottom of the conduction band located just below the Fermi level.

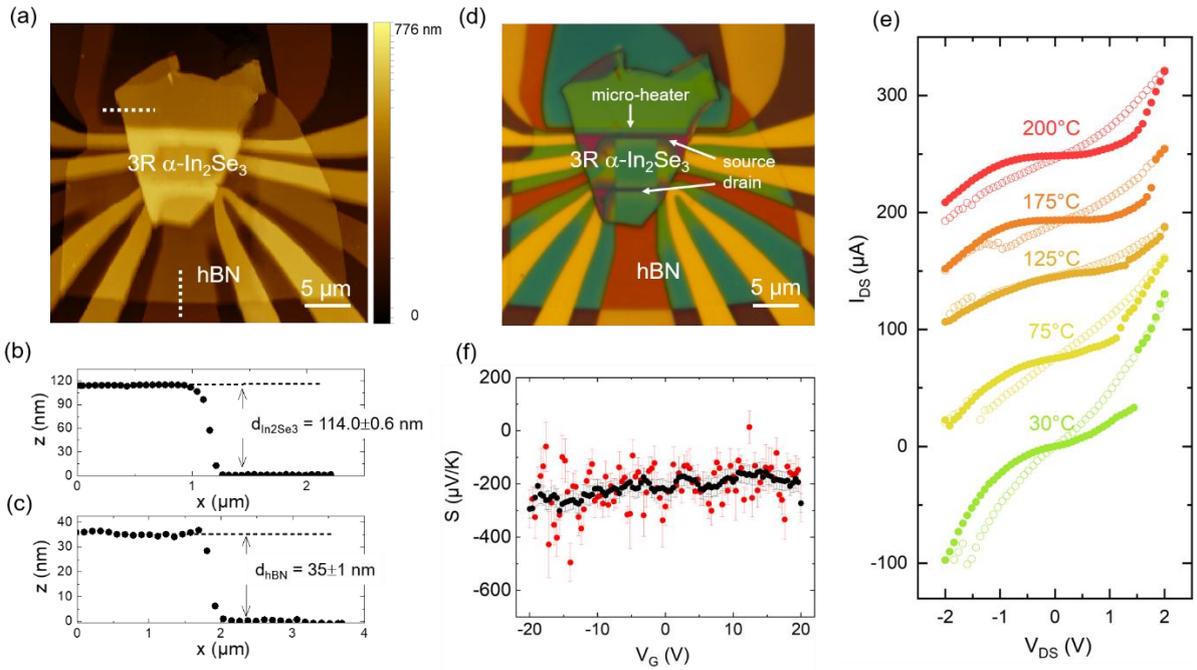

**Figure 5:** (a) AFM image of a representative 3R $\alpha$-In$_2$Se$_3$ based FET. Thicknesses estimation of the 3R $\alpha$-In$_2$Se$_3$ (b) and hBN (c) layers: line-cuts of the flakes indicated by the dotted line in (a), the extracted thicknesses are d$_{In2Se3}$ = 114.0 ± 0.6 nm and d$_{hBN}$ = 35 ± 1 nm. d) Optical image of device in (a). (e) IV characteristics of the 3R $\alpha$-In$_2$Se$_3$-based device in (a) at different temperatures. Measurements are acquired by sweeping the voltage bias from negative to positive values (filled dots) and from positive to negative ones (open dots). (f) Seebeck coefficient measured as a function of the gate voltage of another analogous device, row data are in red and smoothed data in black.

In this work, we have conducted an extensive study exploring the band structure and transport properties of 3R α-In$_2$Se$_3$, demonstrating its remarkable electronic properties. The electronic band structure was studied using both ARPES and density functional theory, highlighting a Mexican-hat band configuration, relevant for novel applications in multiple domains. We performed a detailed analysis of the top valence band, which exhibits an inversion of the effective mass sign near the zone center with a local minimum having depth of 140 ± 10 meV. Also, we have unveiled the occurrence of a highly metallic states at the surface (2DEG) of vacuum-cleaved 3R α-In$_2$Se$_3$ single crystals. This study provides experimental access to robust 2DEGs based on ferroelectric materials, which are scarce in literature. It is of pivotal interest to continue this study by correlating the 2DEG properties and the In$_2$Se$_3$ particular phase.

**Materials and methods**

**Experiments.** The micro-Raman measurements were conducted at room temperature, using a commercial confocal Horiba micro-Raman microscope with a ×100 objective and a 532 / 633 nm laser excitation. XPS/ARPES experiments were performed at the CASSIOPEE beamline of the SOLEIL synchrotron light source at T = 80K. The CASSIOPEE beamline is equipped with a Scienta R4000 hemispherical electron analyzer and the incident photon beam was focused into a <100 $\mu$m spot on the sample surface with linear horizontal polarization. High-quality samples from the "HQ Graphene" company were cleaved at room temperature with scotch tape at a base pressure better than $5 \times 10^{-9}$ mbar. The experiment was performed at room temperature with an energy resolution better than 15 meV.

**Computational details.** The theoretical calculations were carried out by using first principles calculations based on density functional theory (DFT) by means of the Quantum ESPRESSO suite [38]. We used a fully relativistic pseudopotential and non-collinear simulations to consider the spin-orbit interaction and adopted norm conserving pseudopotentials and we used the hybrid Heyd−Scuseria−Ernzerhof (HSE) functional to describe the [23] exchange-correlation term. Van der Waals interactions were considered in the calculations with the semi-empirical Grimme-D3 correction. The self-consistent solution was obtained by adopting a 8x8x8 Monkhorst-Pack grid and a cutoff energy of 80 Ry. Cell parameters and atomic positions were relaxed according to a convergence threshold for forces and energy of $10^{-4}$ and $10^{-8}$ (a.u.), respectively.

**Data Availability statements:** data available from the authors upon reasonable request.

**ACKNOWLEDGMENTS:** We acknowledge the financial support by Optitaste (ANR21-CE24-0002), DEEP2D (ANR-22-CE09-0013), MixDferro (ANR-21-CE09- 0029), and FastNano (ANR-22-PEXD-0006), as well as the French technological network RENATECH.

APPENDIX

Figure S1 displays large scale band structure calculations obtained in the HSE approximation (panel (a)) for $k_z = 0$ (dark red) and $k_z = \pi/c$ (light red), as well as corresponding DOS obtained at the center of the BZ (panel (b)). It shows the negligible dispersion of the VB and at the contrary the strong dispersion of the CB as a function of $k_z$. The 1.3 eV band gap is reported between the maximum of the VB and the minimum of the CB at $k_z = 0$ and at the center of the BZ.

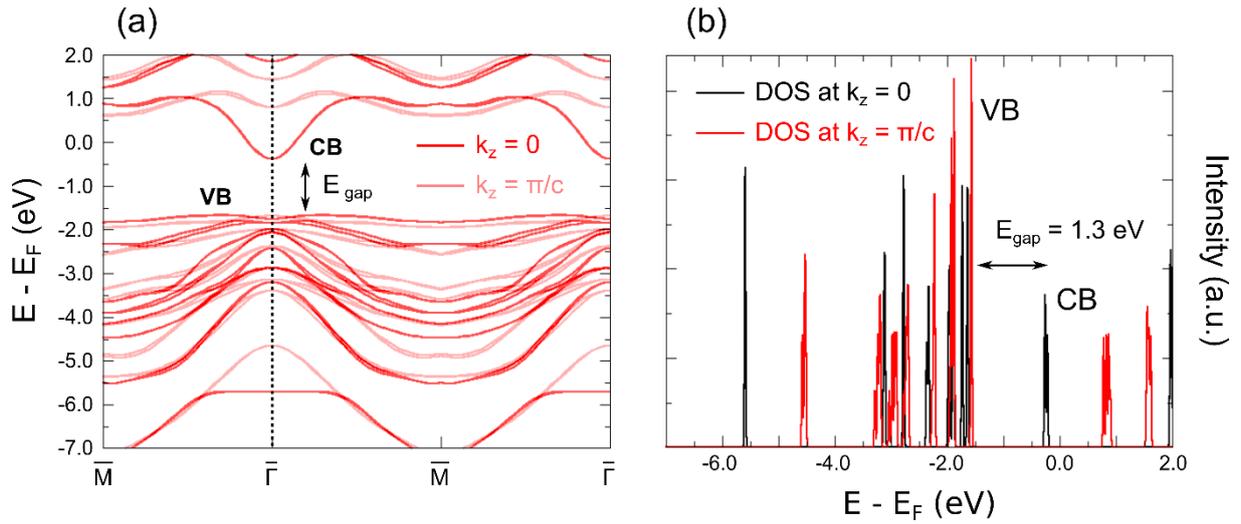

**Figure S1:** (a) Large scale DFT calculations along the $\bar{M} - \bar{\Gamma} - \bar{M} - \bar{\Gamma}$ high-symmetry direction in the HSE approximation for two distinct values of $k_z$ in the 3D BZ, namely $k_z = 0$ (dark red) and $k_z = \pi/c$ (light red). (b) Corresponding DOS extracted at the center of the 2D BZ (black dashed lines in panel (a)) showing a 1.3 eV band gap between the maximum of the VB and the minimum of the CB at $k_z = 0$.

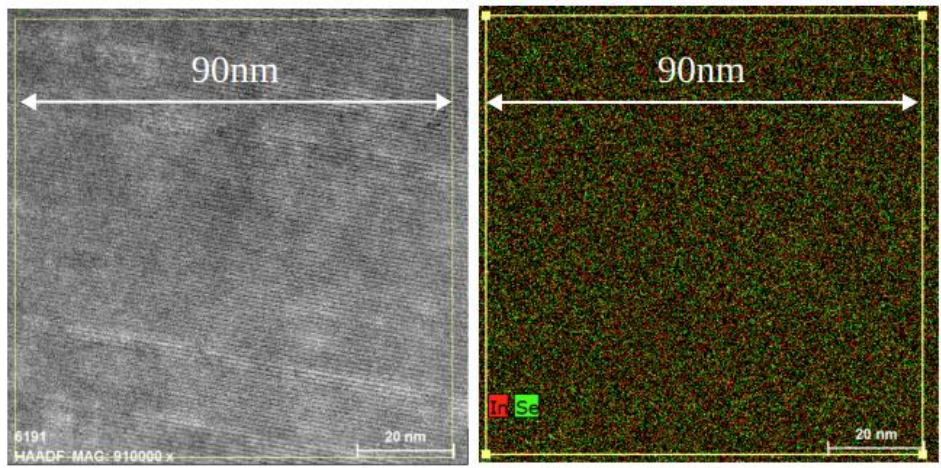

**Figure S2:** EDX elemental maps showing the spatial distribution of In and Se, respectively.